\documentclass[preprint,showpacs,preprintnumbers,amsmath,amssymb,aps,pra]{revtex4}
\usepackage{amssymb,amsmath,graphics,epsfig,amssymb,rotating}
\usepackage{color}
\usepackage{dcolumn}% Align table columns on decimal point%\usepackage{multicol}
\usepackage{bm}% bold math
\usepackage{latexsym}% Include special symbols\
\usepackage{amssymb,amsmath}
\usepackage{color}
\usepackage{subfigure}

\usepackage{titlesec}
\usepackage{wrapfig}
\usepackage{lipsum}

\begin{document}

\title{{\bf Evidence of quadrupole-quadrupole interactions in ultracold gases}}

\author {Jianing Han \footnote[1]{Email address: jhan@southalabama.edu}, Juliet Michell, and Morgan Umstead}
\affiliation{Department of Physics, University of South Alabama, Mobile, Alabama 36688.}

\date{\today}
\begin{abstract}
Van der Waals interactions are interactions between dipoles. Similarly, quadrupole-quadrupole interactions are interactions between quadrupoles. In this article, we focus on the interactions between two dipoles or two quadrupoles. Classically, we treat one Rydberg atom as a dipole; an outer excited electron and an ion core are the two poles of a dipole. Quantum mechanically, we consider Rydberg transition dipoles. Therefore, dipole-dipole interactions are the interactions between two Rydberg atoms. Rydberg atoms have quadrupole components; consequently, the interactions between two Rydberg atoms have quadrupole-quadrupole interaction components. In this article, we examine the dipole-dipole and quadrupole-quadrupole contribution to the interactions between ultracold Rydberg atoms. It is shown that the evidence of quadrupole-blockade has been observed, which is essential for fabricating more compact quantum computers, quantum electronics, as well as quantum sensing.   
\label{abst} 
\end{abstract}

\maketitle
\section{Introduction}
Interactions between charges, or Coulomb interactions, govern three phases of matter in physics as well as other disciplines, such as chemistry, chemical engineering, etc. For example, the chemical bonds, van der Waals bonds, covalent bonds, and ionic and metallic bonds, are based on Coulomb interactions between multiple charges or different orders of multipole-multipole interactions. One type of multipole-multipole interaction is called van der Waals interaction. Van der Waals interactions are second-order dipole-dipole interactions. In this article, van der Waals interactions and higher-order multipole-multipole interactions in excited states are investigated.

Multipole-multipole interactions are related to density broadening. Density broadening is very common in gasses, and those are partially caused by multipole-multipole interactions. The primary contribution to the broadening in longer ranges is the van der Waals interactions. Van der Waals interactions have been studied for over a century. The recent development in laser cooling and trapping  \cite{Wineland, Neuhauser, William, Chu, Lett, Ketterle, Cornell} made it possible to observe such interactions in ultracold gasses \cite{Haroche0, Faoro}. The primary focus of this article is the quadrupole-quadrupole interactions. 

Here is a brief introduction of dipole-dipole and quadrupole-quadrupole interactions. One type of very common dipole-dipole interaction is van der Waals interactions, or the second-order dipole-dipole interactions, which are proportional to $\frac{1}{R^6}$. On-resonance dipole-dipole interactions are proportional to $\frac{1}{R^3}$, which involves two dipole-coupled states. Due to symmetry, on-resonance dipole-dipole interactions can not exist in the same state. However, on-resonance quadrupole-quadrupole interactions can exist in the same state. On-resonance quadrupole-quadrupole interactions are proportional to $\frac{1}{R^5}$. For example, 20d20d states don't have on-resonance dipole-dipole interactions since 20d can not couple with 20d through dipole interactions; however, those states do have on-resonance quadrupole-quadrupole interactions. The second-order quadrupole-quadrupole interactions are proportional to $\frac{1}{R^{10}}$. The second-order quadrupole-quadrupole interactions are small at longer distances compared to the on-resonance quadrupole-quadrupole interactions. However, those interactions can cause asymmetric broadening in the spectra. In this experiment, we study the first-order quadrupole-quadrupole interactions in highly excited atoms or Rydberg atoms. 

This study has many potential applications. For example, dipole-dipole interactions have been proposed to be used as quantum gates \cite{Martin,Haroche,Tong,Singer,Liebish,Jaksch,Lukin,Pierre,Walker1,Heidemann}. Dipole-dipole interactions dominate at lower densities. However, in order to create quantum gates at higher densities or create more compact quantum gates, higher-order multipole-multipole interactions need to be considered. In this experiment, we examine the possibility of using the quadrupole-quadrupole interactions to create quantum gates.

\section{Theory}
%We considered the on-resonance dipole-dipole interactions between 5p and 34d states. 
The Hamiltonian of the system considered is 
\begin{equation}
\begin{split}
H=-\frac{\hslash^2}{2m_1}\nabla^2_1-\frac{\hslash^2}{2m_2}\nabla^2_2-\frac{e^2}{4\pi \epsilon_0}(\frac{1}{r_1}+\frac{1}{r_2}) +V_{dd}+V_{qq},
\end{split}
\label{Hamiltonian}
\end{equation}
where $m_1$ and $m_2$ are the reduced masses of the two electrons. $r_1$ and $r_2$ are the atomic radii. $\epsilon_0$ is the permittivity of free space. $V_{dd}$ and $V_{qq}$ are dipole-dipole and quadrupole-quadrupole interactions respectively, and all higher-order multipole-multipole interactions are ignored in this calculation. The dipole-quadrupole interactions are not included in this calculation. The reason is that the first-order dipole-quadrupole interaction, which is proportional to $1/R^4$, vanishes in the case considered. The second-order dipole-quadrupole interactions are proportional to $1/R^8$, which is a higher-order interaction than the first-order quadrupole-quadrupole interactions and the second-order dipole-dipole interactions, or van der Waals interactions.

In this article, we focus on the highly excited states, Rydberg states \cite{Gallagher11, Dunning}. The on-resonance dipole-dipole interaction potential energy is 
\begin{equation}
\begin{split}
V_{dd}=-\frac{p_1p_2}{4\pi \epsilon_0}\frac{(C_{1,-1}^1C_{1,1}^2+C_{1,1}^1C_{1,-1}^2+2C_{1,0}^1C_{1,0}^2)}{R^3},
\end{split}
\label{Vqq}
\end{equation}
where $p_1$ is the dipole moment of the first atom, and $p_2$ is the dipole moment of the second atom. $C_{k,q}$ is a spherical tensor \cite{Edmonds}. %$e$ is the electron charge. $r_1$ and $r_2$ are the radii of atom 1 and atom 2 respectively. 
$R$ is the distance between the two quadrupoles. On-resonance dipole-dipole interactions \cite{WHGallagher2}, which are proportional to $\frac{1}{R^3}$, may also cause broadening. However, the on-resonance dipole-dipole coupling is caused by the dipole between the 5p atoms and the 34d atoms in the case studied, and the size of this broadening is negligible due to the big size difference between the 5p atoms and the Rydberg atoms, 34d atoms. 

The van der Waals interaction, $V_{vdw}$, is
\begin{equation}
\begin{split}
V_{vdw}=\frac{V_{dd}^2}{\Delta},
\end{split}
\label{Vvdw}
\end{equation}
where $\Delta$ is the energy difference between the two energy levels considered. Since $V_{dd}$ is proportional to $\frac{1}{R^3}$ as shown in Eq. \eqref{Vvdw} and van der Waals interactions are proportional to $V_{dd}^2$, which leads to the $\frac{1}{R^6}$ dependence of van der Waals interactions. %Few-body interactions are the sums of all the pair-wise interactions \cite{}. 
The closest van der Waals coupled state in this case is 33f$_{5/2}$35p$_{3/2}$, which is 4.6 GHz above the 34d$_{5/2}$34d$_{5/2}$ state and will cause attractive interactions on the 34d$_{5/2}$34d$_{5/2}$ state. This can be partially explained by the experimental data as shown in the result and discussion session.

The quadrupole-quadrupole interaction potential energy is the following:
\begin{equation}
\begin{split}
V_{qq}&=\frac{Q_1Q_2}{4\pi \epsilon _0R^5}\{C_{2,2}^{1}C_{2,-2}^{2}+C_{2,-2}^{1}C_{2,2}^{2}+4[C_{2,1}^{1}C_{2,-1}^{2}+C_{2,-1}^{1}C_{2,1}^{2}] +6C_{2,0}^{1}C_{2,0}^{2}\},
\end{split}
\label{Vqq}
\end{equation}
%notes page 131.
where $Q_1=er_1^2$ and $Q_2=er_2^2$ are the quadrupole moments of atom 1 and atom 2 respectively. On-resonance quadrupole-quadrupole interactions are proportional to $\frac{1}{R^5}$. In addition, the second-order quadrupole-quadrupole interaction, $V_{qq2}$, are also included in this calculation. 
\begin{equation}
\begin{split}
V_{qq2}=\frac{V_{qq}^2}{\Delta},
\end{split}
\label{Vqq2}
\end{equation}
Since the on-resonance quadrupole-quadrupole interactions, $V_{qq}$, are proportional to $\frac{1}{R^5}$, and the second-order quadrupole-quadrupole interactions are proportional to $V_{qq}^2$, which results in the $\frac{1}{R^{10}}$ dependence of the second-order quadrupole-quadrupole interactions, or $V_{qq2}\propto \frac{1}{R^{10}}$. The effect of the second-order quadrupole-quadrupole interactions is negligible for the case considered. The calculation was done by diagonalizing the matrix of the Hamiltonian with the quadrupole-quadrupole potential energy. The details of the calculation are described in the result and discussion session.

\section{Experiment}
In this experiment, we first create a $^{85}$Rb Magneto-Optical Trap (MOT) \cite{WHGallagher, Lett, Cornell}. The density of the MOT is about $10^{10}$ cm$^{-3}$. The temperature of the MOT is about 300 $\mu$K. %The total number of atoms is about 10^{7}. 
We then excite the atoms to a highly excited state, or a Rydberg state \cite{Gallagher11}, at time zero. In this experiment, we use a pulsed dye laser to do the excitation, and we use a wavemeter to monitor the laser output and make sure that the spontaneous emission is suppressed. The pump pulse width from the Nd:YAG laser is about 5-6 ns. The line width of the dye laser is limited by the Fourier transform of the dye laser pulse%, which will be discussed in the result and discussion section
. We then ionize the atoms using selective field ionization (SFI) \cite{GallagherFI}. The rising time of the field ionization pulse is about 5 $\mu$s. Different Rydberg states will be ionized at different times. The peak voltage is about 1850 V. We detect the result ions using a pair of Microchannel plates (MCPs). We then collect the ions and display the signal on an oscilloscope. Each peak on the oscilloscope corresponds to one Rydberg state. The oscilloscope signal is exported to a computer and analyzed. For example, we integrate one peak on the oscilloscope to get the number of Rydberg atoms in that particular state. Due to the fact that the field ionization pulse is added after the Rydberg excitation pulse, there is no Stark effect associated with the Rydberg signal collected. Such Rydberg atom detection techniques can be enhanced by combining them with imaging techniques. Recently, Rydberg atoms have been detected by imaging \cite{Browaeys}. 

\begin{figure}[tpb]
\centering
\includegraphics[width=9.3 cm, angle= 0, height=9.3 cm]{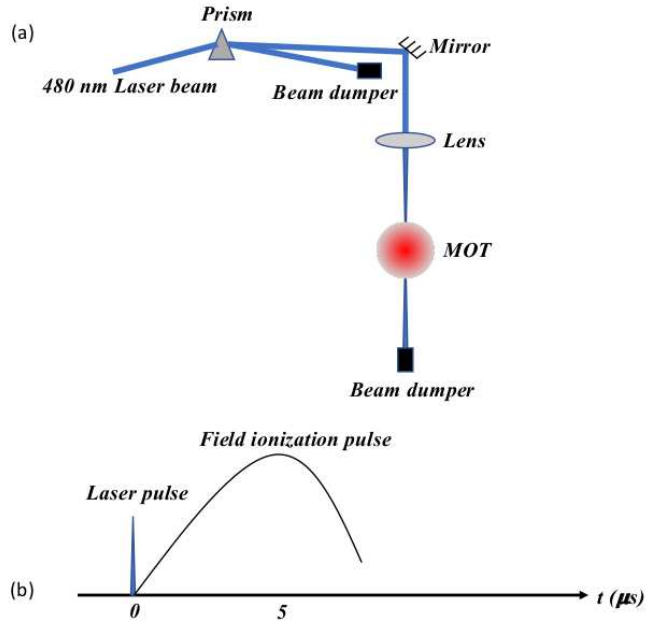}
\caption{  (a) The schematic plot of the laser excitation. (b) The timing of this experiment.}
 \label{schematic}
\end{figure}

The experimental setup and the timing of this experiment are shown in Fig. \ref{schematic}. A prism is added at the output of the dye laser, so the fluorescence is scattered off. The laser beam shown in Fig. \ref{schematic}(a) comes out of the laser and first passes through a prism. The laser beam is split into two beams, the spontaneous emission, 480 nm, and the coherent radiation from the cavity, 481.3 nm. The spontaneous emission is refracted and blocked by a beam dumper after the prism. The coherent laser beam, 481.3 nm, is then reflected by a mirror and then focused on the MOT inside a vacuum chamber. The rest of the laser beam is blocked by a beam dumper. The timing of this experiment is shown in Fig. \ref{schematic}(b). The atoms are excited to the 34d$_{5/2}$ state at time t=0 $\mu$s followed by a field ionization pulse, which ionizes the excited atoms. 

The frequency of the laser is tuned by tuning one of the gratings of the dye laser. As shown in previous references \cite{Han2017}, broadband lasers can be used to suppress dipole-blockade, which will, in turn, allow the close neighboring atom interactions. This experiment directly tests those theories. 

\section{Result and discussion}

\begin{figure}[tpb]
\centering
\includegraphics[width=6 cm, angle= 0, height=6 cm]{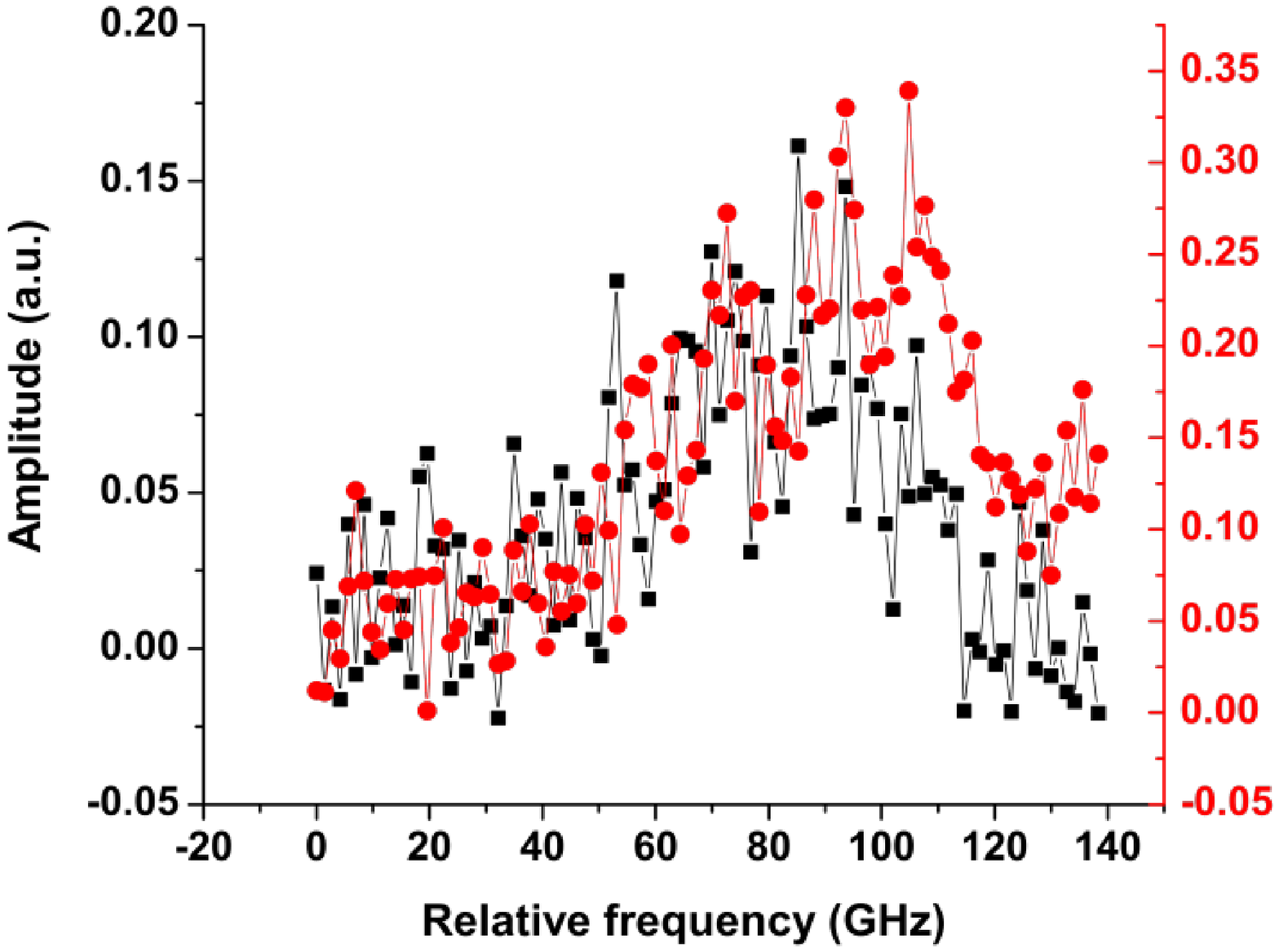}
\caption{  Two laser scans are taken at two different densities 0.89$\times10^{9}$ cm$^{-3}$ (-$\blacksquare$-) and 5.01$\times10^{9}$ cm$^{-3}$ (\textcolor{red}{-$\bullet$-}). %The actual density may be higher due to the diabatic ionization of the higher angular momentum state mixing. The greater density will lead to a greater error due to the greater mixing probability. 
}
 \label{Experimental_data}
\end{figure}

Fig. \ref{Experimental_data} shows two scans taken by scanning the laser frequency. Each data point is the average of ten laser shots. The frequency is characterized by measuring the energy difference between the 32$d_{5/2}$ and 33s, which are separated by 186.57567 GHz. The lower density scan is in black, and the higher density data is in red, as shown in Fig. \ref{Experimental_data}. Two features are clearly visible. First, the higher density data is broadened. Second, the excitation is suppressed at higher densities by dividing the peak height by the density, which shows the feasibility of using such interactions for quantum gates. More specifically, it is quadrupole-blockade similar to dipole-blockade, and in most cases, the suppression is caused by two or more types of multipole-multipole interactions. Only two laser scans are shown here. The high-density suppression and line-broadening features are repeatable. 

\begin{figure}[tpb]
\centering
\includegraphics[width=6 cm, angle= 0, height=6 cm]{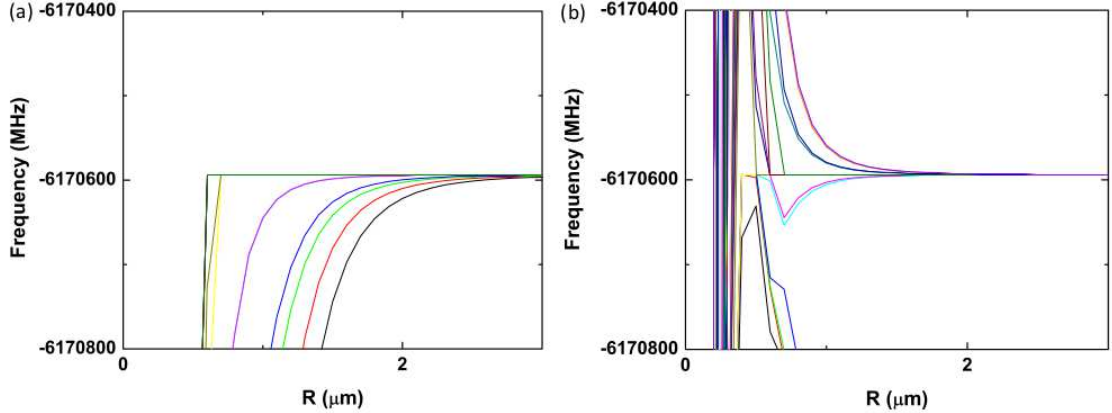}
\caption{  (a) The van der Waals interactions between 34d$_{5/2}$34d$_{5/2}$ and neighboring states. (b) The quadrupole-quadrupole interactions between 34d$_{5/2}$34d$_{5/2}$ and the state itself and neighboring states.}
 \label{Calculation}
\end{figure}

Fig. \ref{Calculation} shows the calculations of the energy levels calculated for the two-body case. Fig. \ref{Calculation}(a) shows the van der Waals interactions in the 34d$_{5/2}$34d$_{5/2}$ state, and Fig. \ref{Calculation}(b) shows the quadrupole-quadrupole interactions in the same state. In both calculations, two-body interactions are considered. Since in the two-body calculations, the $M$ is conserved, where $M$ is the projection of total angular momentum in the z-axis, only M=0 states are considered. Three principal quantum numbers, 33, 34, and 35, are included in this calculation, and a total of 2700 matrix elements are taken into account. If van der Waals interactions are the only broadening mechanism, then the broadening should be to the lower frequency end. However, the experimental data shows the broadening to both higher and lower frequencies and slightly more on the higher frequency side, which corresponds to the broadening caused by the quadrupole-quadrupole interactions. The fact that the higher density data shows more broadening in the higher frequency side indicates that this is caused by short-range interactions, such as quadrupole-quadrupole interactions, which is consistent with the calculation shown in Fig. \ref{Calculation}. However, the size of the broadening from the calculation is much less than the experimental observation, which indicates the multi-body nature of this problem. In other words, few-body or many-body interactions need to be considered in order to explain the experimental data fully. The potential well in Fig. \ref{Calculation}(b) also shows the possibility of creating ultracold quadrupole-quadrupole molecules. Such potentials are caused by avoided crossings \cite{Farooqi, Overstreet, Li}.

\begin{figure}[tpb]
\centering
\includegraphics[width=6 cm, angle= 0, height=6 cm]{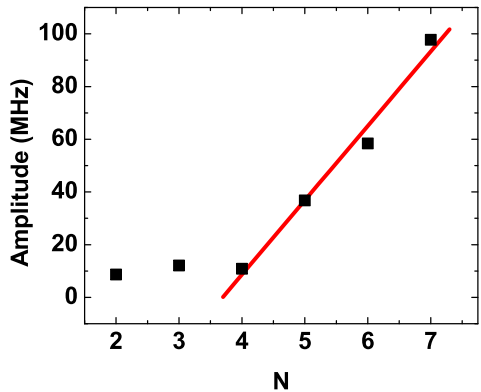}
\caption{  The frequency shift at one micrometer vs. the total number of atoms, N, considered.}
 \label{Scaling_linear}
\end{figure}

To give an estimation of the lowest number needed to be considered in this calculation, we studied the few-body interactions. For simplicity, we only consider the 34d$_{5/2}$ state, and this calculation is done for a one-dimensional case. Fig. \ref{Scaling_linear} shows the maximum frequency shift at 1 $\mu$m, the nearest neighbor spacing between the few bodies, for different numbers of atoms. It is expected that as the number of bodies increases, the amount of frequency shift will increase. Specifically, the maximum frequency shift is proportional to the number of bodies. A linear fit is applied to the four to seven bodies' data, and it agreed well. %It is anticipated that the slope of the curve will decrease as N approaches infinity. 
%From this linear fit, it is estimated in order to have a 10 GHz frequency shift, 350 atoms need to be considered. In reality, the number of interacting bodies should be much greater than this number. %The maximum frequency shift calculated at one micrometer, the nearest neighbor spacing between the few bodies, vs. the number of interacting bodies. are show

In this article, we choose a state so that the near-resonance interactions are negligible to study the quadrupole-quadrupole interactions. The on-resonance dipole-dipole interaction is the interaction between 5p$_{3/2}$ and 34d$_{5/2}$. The transition dipole moment between 5p$_{3/2}$ and 34d$_{5/2}$ is about 0.06 ea$_0$, which is about 10$^{-4}$ of the dipole moment between the $34d_{5/2}$ and its dipole-transition allowed neighboring Rydberg states. We have calculated two-body on-resonance dipole-dipole interactions. Since M, the projection of the sum of the angular momentum on z-axis, is conserved, we only considered one M state, the M=0 state, for simplicity. The results of all other M states are similar to the state reported here. All those calculations indicate that the on-resonance dipole-dipole interactions between 5p$_{3/2}$ and 34d$_{5/2}$ are negligible in the problem considered here. 

It has been shown that as the density increases, the offset of the whole spectrum moves up, which indicates that there is a broadband under the atomic transition, which is similar to the energy band in a solid. However, no band structure has been observed in this experiment. This is consistent with the previous prediction that a cold Rydberg gas is similar to an amorphous material \cite{WHGallagher2}.

Charge-multipole interactions may play a role when ions are present or at very high densities when plasma is created. In this experiment, we tried to reduce those effects. For example, we monitor the laser to reduce the spontaneous emission, which may produce ions. In addition, we detect the interactions right after the atoms are excited to Rydberg states to reduce the probability of evolving into ultracold plasma.

\section{Conclusion}
It is shown that the broadening of the 34d$_{5/2}$34d$_{5/2}$ states is partially caused by the van der Waals interactions and partially caused by the quadrupole-quadrupole interactions. In addition, it has been shown that the quadrupole-quadrupole interactions show repulsive interactions. Moreover, excitation suppression has been observed, or the evidence of partial quadrupole blockade has been observed. The experimental data can be qualitatively explained by the two-body multipole-multipole interactions. More quantitative measurements are underway. In order to fully explain the experimental data, few-body or many-body multipole-multipole interactions are required. Furthermore, the calculations show that it is possible to create quadrupole-quadrupole coupled molecules caused by the avoided crossings.

\section{Acknowledgement}  
    It is a pleasure to acknowledge the support from the Air Force Office of Scientific Research (AFOSR), a previous award from the Army Research Office (ARO), and the University of South Alabama Faculty Development Council (USAFDC).

\end{document}